\documentclass[preprint,prd,amsmath,amssymb,amsthm,nofootinbib]{revtex4}
\usepackage{xcolor}
\usepackage[mathscr]{euscript}
\usepackage{bm}% bold math
\usepackage{graphicx}% Include figure files
\usepackage[caption=false]{subfig}
\usepackage[colorlinks=true,linkcolor=red]{hyperref}
\usepackage{ulem}
\usepackage{soul}              % for crossing things out
\newcommand{\bea}{\begin{eqnarray}}
\newcommand{\eea}{\end{eqnarray}}
\newcommand{\beq}{\begin{equation}}
\newcommand{\eeq}{\end{equation}}

\begin{document}
\title{ Quantum parameter estimation for detectors in constantly accelerated motion }
\author{Han Wang$^{1}$, Jialin Zhang$^{1,2}$~\footnote{Corresponding author. jialinzhang@hunnu.edu.cn} and Hongwei Yu$^{1,2}$~\footnote{Corresponding author. hwyu@hunnu.edu.cn}}
\affiliation{
$^1${\small{Department of Physics, Key Laboratory of Low-Dimensional Quantum Structures and Quantum Control of Ministry of Education, Hunan Research Center of the Basic Discipline for Quantum Effects and Quantum Technologies, Hunan Normal University, 36 Lushan Rd., Changsha, Hunan 410081, China}}\\
$^2${\small{Institute of Interdisciplinary Studies, Hunan Normal University, 36 Lushan Rd., Changsha, Hunan 410081, China}}}

\begin{abstract}
We analyze quantum parameter estimation by studying the dynamics of the quantum Fisher information (QFI) for two classes of parameters, acceleration and initial-state weight, in an Unruh-DeWitt detector undergoing four distinct noninertial motions: linear, cusped, catenary, and circular trajectories respectively. We assume that the detector is initialized in a pure superposition state with a weight parameter
 $\theta$  characterizing the probability of the detector occupying each state. Our results reveal that, over long evolution times, the QFI for the acceleration parameter converges to a nonnegative asymptotic value that depends sensitively on the trajectory, whereas the QFI for the weight parameter decays to zero as the system thermalizes. Importantly, for sufficiently large accelerations, one can attain the optimal precision in estimating the acceleration parameter within a finite interaction time, eliminating the need for infinitely long measurements. Comparing trajectories, we find that for small accelerations (relative to the detector's energy gap), linear motion yields the highest QFI for
 $\theta$, while for large accelerations, circular motion becomes optimal for estimating
 $\theta$. By contrast, circular motion offers the best precision for estimating acceleration itself in both the small- and large-acceleration regimes (the latter only at very long times). These contrasting behaviors of QFI across trajectories suggest a novel metrological protocol for inferring the underlying noninertial motion of a quantum probe.
\end{abstract}

\maketitle

\section{Introduction}

Quantum metrology, an interdisciplinary field dedicated to enhancing measurement precision through quantum properties, has become a cornerstone of modern quantum information science~\cite{Giovannetti:2004,Giovannetti:2006,Paris:2009,Giovannetti:2011}. A fundamental challenge in this field arises from the fact that the physical parameter of interest often lacks a direct quantum observable, leading to inherent measurement inaccuracies and statistical uncertainties. To address this, quantum estimation theory~\cite{Helstrom:1976,Holevo:1982} has emerged as a crucial subfield, with quantum Fisher information (QFI)~\cite{Braunstein:1994} playing a central role.

Mathematically, QFI quantifies the sensitivity of a quantum state to variations in the parameter of interest, forming the foundation of quantum estimation theory~\cite{Paris:2009}. Its theoretical significance is underscored by the Cram\'{e}r-Rao inequality~\cite{Cramer:1946}, which establishes QFI as the fundamental bound on estimation precision. This relationship implies that higher QFI directly correlates with enhanced measurement accuracy in quantum metrology tasks.

QFI has found widespread applications across various quantum technologies, including quantum frequency standards~\cite{Bollinger:1996}, optimal quantum clocks~\cite{Buzek:1999}, quantum phase estimation~\cite{Giovannetti:2006,Dorner:2009}, non-Markovianity characterization~\cite{Suncp:2010}, uncertainty relations~\cite{Luo:2003,Gibilisco:2007}, and entanglement detection~\cite{Li:2013}, among others. With the rapid advancement of quantum information technologies, relativistic effects in quantum metrology are becoming increasingly relevant at a practical level.

Recently, significant interest has emerged in exploring the interplay between quantum metrology and relativistic phenomena. This includes applications in estimating the Unruh-Hawking effects~\cite{Ahmadi:2013,Aspachs:2010,Ahmadi:2014,Peters:1999,Hosler:2013,Yao:2014,Wang:2014,Hao:2015,Tian:2015,Zhao:2021,Abd-Rabbou:2022,Du:2021,Patterson:2023}, gravitational redshift~\cite{Muller:2010}, and parameters of Schwarzschild spacetime~\cite{Bruschi:2014}. Additionally, QFI has been applied to the study of expansion rates in the Robertson-Walker universe~\cite{Wang:2015}, quantum metrology with indefinite causal order~\cite{zhao:2020}, and the interaction between Bose-Einstein condensates and gravitational waves~\cite{Sabin:2014,Schutzhold:2018,Robbins:2019}. These developments underscore the growing intersection between quantum information science and relativistic physics, paving the way for novel quantum technologies that leverage fundamental principles from both fields.

One intriguing example of the interplay between quantum metrology and relativistic phenomena is parameter estimation for uniformly accelerated observers. Such observers perceive the Minkowski vacuum as a thermal bath, giving rise to the Unruh effect, one of the most striking predictions of relativistic quantum field theory. As a fundamental counterpart to Hawking radiation of black holes and Gibbons-Hawking effect in curved spacetime, the Unruh effect provides crucial insights into the nature of black hole evaporation and cosmological horizon thermality. Recent theoretical advancements in quantum estimation techniques suggest that measurements of the Unruh effect may be feasible at experimentally accessible accelerations~\cite{Aspachs:2010,Ahmadi:2014}.

The Unruh-DeWitt detector~\cite{Unruh:1976,DeWitt:1979}, originally introduced to illustrate the Unruh effect, has proven to be a valuable quantum probe for characterizing relativistic effects via open quantum system dynamics and quantum metrology~\cite{Suncp:2010,Jin:2015,Tian:2015,Wang:2015,Du:2021,Patterson:2023}. In particular, by analyzing the evolution of QFI for accelerated detectors, Ref.~\cite{Tian:2015} explored quantum estimation of the Unruh effect, arguing that optimal precision in estimating acceleration can be achieved when the detector evolves over a sufficiently long duration. However, this conclusion primarily holds for linear acceleration with moderate magnitudes. Its validity for large accelerations and its applicability to other constant acceleration scenarios in Minkowski spacetime remain unexplored.

It is therefore crucial to investigate whether alternative acceleration profiles can yield better performance in quantum estimation of Unruh-like effects, particularly when employing adaptive strategies or feedback mechanisms to approach the ultimate precision bound on acceleration estimation. Additionally, an Unruh-DeWitt detector should inherently be considered as an open quantum system coupled to a fluctuating field environment. The interaction with this environment introduces noise into quantum measurements, which in turn affects the precision limits of estimating not only acceleration but also other relevant parameters~\cite{Huelga:1997,Rosenkranz:2009,cpsun:2011,Jin:2015}.
It is important to emphasize that Unruh-DeWitt detectors following different accelerated trajectories may exhibit distinct dynamical behaviors and capture different aspects of Unruh-like thermalization phenomena, such as varying power spectra of Rindler noise and trajectory-dependent effective temperatures. In this sense, different acceleration profiles can be viewed as effectively corresponding to distinct environmental conditions, which can significantly impact the QFI associated with parameter estimation. By analyzing QFI for parameters beyond acceleration, one can assess the extent of noise induced by various acceleration profiles and its impact on measurement precision.

In this paper, we employ quantum metrology techniques,  specifically QFI and  the Unruh-DeWitt detector to comprehensively address several fundamental yet previously overlooked questions: (1) under what conditions can acceleration be estimated with the highest precision?  (2) which acceleration scenarios optimize the precision of acceleration estimation? and (3) which acceleration profiles provide maximal precision for estimating parameters other than acceleration, given the noise introduced by the motion itself? By systematically addressing these questions, we aim to deepen our understanding of relativistic quantum metrology and its implications for high-precision measurements in accelerating reference frames.

The paper is organized as follows. In Sect. II, we begin by briefly reviewing the basic formalism of the Unruh-DeWitt detector model within the framework of open quantum system theory, as well as QFI in the context of quantum metrology. In Sec. III, after introducing various common accelerated  motions, we proceed to calculate the QFI for both the acceleration parameter and the weight parameter  characterizing the state of the accelerated detector. Both the analytical and numerical results collectively reveal  the distinct characteristics of quantum metrology in different acceleration scenarios. Finally, we conclude this paper in Sec. IV.

%%%%%%%%%%%%%%%%%%%%%%%%%%%%%%%%%%%%%%

\section{ The Unruh-DeWitt detector model and quantum Fisher information }
\label{sec2}
\subsection{The Unruh-DeWitt detector model}
In general, a probe quantum system, on which measurements are performed, does not exist in isolation but rather functions as an open quantum system. Consequently, its temporal evolution is typically nonunitary, as interactions with an external environment can significantly influence its dynamics. These environmental effects introduce system-dependent noise and decoherence, leading to environment-dependent properties in quantum metrology.

Without the loss of generality,  we employ the Unruh-DeWitt detector, which is modeled as  a two-level  quantum system, as the probe system.  The two energy levels  of the detector are
respectively denoted as $|0\rangle_{D}$ and $|1\rangle_{D}$ with an
energy gap $\Omega$. Thus, the total
Hamiltonian of the combined system of the Unruh-DeWitt detector plus
quantum fields can be written in the form
\begin{equation}
 H=H_S+H_F+H_I\;,
\end{equation}
 where
\begin{equation}
H_{S}=\frac{1}{2}\Omega\big(|1\rangle_{D} \langle1|_{D}-|0\rangle_{D}\langle0|_{D}\big)=\frac{1}{2}\Omega\sigma_3\;
 \end{equation}
 is the free Hamiltonian of the detector with $\sigma_{i}~(i=1,2,3)$  representing  Pauli matrices,  $H_{F}=\sum_{\mathbf{k}}\omega_{\mathbf{k}}a_{\mathbf{k}}^{\dagger}a_{\mathbf{k}}$ denotes the Hamiltonian
of  free scalar fields, and $H_I$  represents the interaction
Hamiltonian between the detector and the scalar field $\phi(x)$.
Specifically, for a two-level  system with a trajectory $x(\tau)$
parametrized by its proper time $\tau$,   the interaction
Hamiltonian then can be written as
\begin{equation}
H_I=\lambda({e}^{i\Omega\tau}\sigma^++e^{-i\Omega\tau}\sigma^-)\otimes\phi[x(\tau)]\;,
\end{equation}
where $\lambda$ is the coupling constant,  and $\sigma^{+}=|1\rangle_{D} \langle0|_{D}$ and $\sigma^{-}=|0\rangle_{D} \langle1|_{D}$ denote ladder operators of the detector.

We suppose that  the initial density matrix of the combined  system has the form  of $\rho_{\rm{tot}}(0)= \rho(0)\otimes|0\rangle \langle0|$, where $\rho(0)$ denotes the initial reduced density matrix of the detector, and $|0\rangle$ is the vacuum  state of the  field. Generally,  the evolution of the combined system in the proper time obeys the von Neumann equation
\begin{equation}
\frac{\partial\rho_{\mathrm{tot}}(\tau)}{\partial\tau}=-i[H,\rho_{\mathrm{tot}}(\tau)]\;.
\end{equation}
In the context of open quantum system theory, one can obtain the state of the detector by taking
partial trace over the field degrees of freedom, i.e., $\rho(\tau)=\rm{tr}_\phi[\rho_{\rm{tot}}(\tau)]$.  In the limit of weak coupling,
the reduced density matrix is found to obey an equation in the Kossakowski-Lindblad form~\cite{Gorini:1975,Lindblad:1975,Benatti:2003},
\begin{equation}\label{masterf}
\frac{\partial\rho(\tau)}{\partial\tau}=-i[H_{\mathrm{eff}},\rho(\tau)]+{\mathcal{L}}[\rho(\tau)]\;,
\end{equation}
where  $H_{\mathrm{eff}}$ denotes the effective Hamiltonian of the detector, and ${\mathcal{L}}[\rho]$ is called  the dissipator of the equation.

In general,  the effective Hamiltonian  and  the dissipator  are determined by the Wightman function of fields,
$W\big(x(\tau),x'(\tau')\big)=\langle0|\phi(x)\phi(x')|0\rangle$.  For convenience,  we define  the  Fourier  transforms  of the Wightman function  and  its corresponding Hilbert transform  as follows:
\begin{equation}\label{gomega}
\mathcal{G}(\Omega)=\int_{-\infty}^{\infty}d\Delta \tau e^{i\Omega\Delta\tau} W(\Delta\tau)\;,
\end{equation}
\begin{equation}\label{komega}
\mathcal{K}(\Omega)=\frac{1}{i\pi} \mathrm{PV}\int_{-\infty}^{\infty}d\omega \frac{\mathcal{G}(\omega)}{\omega-\Omega}\;
\end{equation}
with $\mathrm{PV}$ representing the principle value of an integral. %$\Delta\tau=\tau-\tau'$ and .
Then the effective Hamiltonian takes the form
\begin{equation}\label{heff-r}
H_{\rm{eff}}:=\frac{1}{2}\tilde{\Omega}\sigma_3=\frac{1}{2}\Big\{\Omega+\frac{i\lambda^2}{2}\Big[\mathcal{K}(-\Omega)-\mathcal{K}(\Omega)\Big]\Big\}\sigma_3\;
\end{equation}
with $\tilde{\Omega}$ representing the renormalized energy gap that includes the Lamb shift, and
the  dissipator is given by
\begin{equation}
{\mathcal{L}}[\rho]=\frac{1}{2}\sum_{i,j=1}^{3}C_{ij}\left(2\sigma_{j}\rho\sigma_{i}-\sigma_{i}\sigma_{j}\rho-\rho\sigma_{i}\sigma_{j}\right)\;,
\end{equation}
where  the Kossakowski matrix $C_{ij}$ is  defined as
\begin{equation}
C_{ij}=A\delta_{ij}-i B\epsilon_{ijk}\delta_{k3}-A\delta_{i3}\delta_{j3}\;
\end{equation}
with
\begin{equation}\label{ABcs}
A=\frac{\lambda^2}{4}[\mathcal{G}(\Omega)+\mathcal{G}(-\Omega)],\quad
B=\frac{\lambda^2}{4}[\mathcal{G}(\Omega)-\mathcal{G}(-\Omega)]\;.
\end{equation}

For a two-level system, the quantum state  can usually be written  in terms of the Bloch sphere representation
\begin{equation}\label{Blochv}
\rho=\frac{1}{2}(I+\bm{\omega}\cdot\bm{\sigma})\;,
\end{equation}
where $\bm{\omega}:=(\omega_1,\omega_2,\omega_3)$ is just the Bloch vector and $\bm{\sigma}:=(\sigma_1,\sigma_2,\sigma_3)$ .
In principle, one can exactly solve the master equation~(\ref{masterf})  by applying the Bloch sphere representation of quantum state~(\ref{Blochv}).
Suppose the  detector is initially prepared in a pure state
\begin{equation}\label{in-state}
|\psi\rangle_D=\cos\frac{\theta}{2}|1\rangle_{D}+\sin\frac{\theta}{2}|0\rangle_{D}
\end{equation}
 with $\theta$ denoting the weight parameter.
 According to Eqs.~(\ref{masterf}),~(\ref{Blochv}) and~(\ref{in-state}),  the evolution of the Bloch vector can  be solved as follows:
\begin{equation}\label{bloch-re}
\left\{\begin{aligned}&\omega_{1}(\tau)=e^{-2A\tau}\sin\theta\cos\tilde{\Omega}\tau\\
&\omega_{2}(\tau)=e^{-2A\tau}\sin\theta\sin\tilde{\Omega}\tau,\\
&\omega_{3}(\tau)=e^{-4A\tau}\cos\theta-\frac{B}{A}(1-e^{-4A\tau})\;.
\end{aligned} \right.
\end{equation}
It is easy to see that the temporal evolution of  the detector's state is dependent on the environment, as the  parameters $A, B, \tilde{\Omega}$  in Eq.~(\ref{bloch-re}) are associated with external environment.
Note that in the limit of $\tau\rightarrow\infty$,  the Bloch vector becomes $\bm{\omega}=(0,0,-B/A)$, which is independent of the weight parameter of the state.

\subsection{Quantum Fisher information}

QFI plays a crucial role in quantum metrology, serving as a fundamental tool for leveraging quantum resources to achieve higher measurement precision. The parameter estimation problem focuses on determining the value of a nondirectly observable parameter of a quantum system by analyzing a set of measurements performed on directly observable quantities using quantum statistical methods. Given a quantum state parameterized by an unknown parameter $\xi$, its value can be inferred through a set of positive-operator-valued measures (POVMs) applied to a directly observable parameter $x$ of the state.

Since any  measurement data can not be immune to noise,  any
estimation of the parameter $\xi$  will contain some degree of error.  To quantify the precision limit of an estimation, one can introduce  the
Cram\'{e}r-Rao bound , which provides a lower bound on the variance of an
estimator (equal to the mean-squared error for an unbiased
estimator). The Cram\'{e}r-Rao bound is given
by~\cite{Cramer:1946,Helstrom:1969,Holevo:1982}
\begin{equation}
\mathrm{Var}(\xi)\geqslant\frac{1}{N F_C(\xi)}\;,
\end{equation}
where $N$ represents the number of repeated measurements, and $F_C(\xi)$ denotes the classical Fisher information of the parameter of interest,
 \begin{equation}\label{Fxi1}
F_C(\xi)=\int dx~p(x|\xi)\Big[\frac{\partial\ln{p}(x|\xi)}{\partial\xi}\Big]^2=\int dx \frac{1}{p(x|\xi)}\Big[\frac{\partial{p}(x|\xi)}{\partial\xi}\Big]^2\;.
\end{equation}
Here,  $p(x|\xi)$ denotes the conditional probability of  obtaining the value $x$ when the parameter has the value $\xi$.
 The Cram\'{e}r-Rao bound, determined by the Fisher information,  provides a bound on how well the estimation can be performed, i.e., the larger the Fisher information $F_C(\xi)$, the lower the bound on the variance,  thereby allowing for more accurate estimation of the parameter  $\xi$.  By introducing the symmetric logarithmic derivative, $L_\xi$, which is  a self-adjoint operator satisfying the equation $\partial_\xi\rho:=(L_\xi\rho+\rho{L_\xi})/2$, one can further prove that for any quantum measurement
$F_C(\xi)\leq F_Q(\xi):={\rm{tr}}({\rho{L_\xi}^2})={\rm{tr}}(\partial_\xi\rho{L_\xi})$, where $F_Q(\xi)$ is the so-called quantum Fisher information (QFI)~\cite{Braunstein:1994}.
Optimizing over all possible quantum measurements,  an even stronger lower bound emerges in the quantum version of the Cram\'{e}r-Rao theorem,
\begin{equation}
\mathrm{Var}(\xi)\geqslant\frac{1}{N F_Q(\xi)}\;.
\end{equation}

For an orthonormal basis in which any quantum state can be expressed as
$\rho=\sum_{i}p_i|\psi_i\rangle\langle\psi_i|$ with $p_i$  denoting the probability that the quantum system is in state $|\psi_i\rangle$,  the symmetric logarithmic derivative  can, in principle,  be  solved
for the QFI. This leads to the following expression~\cite{Paris:2009,Braunstein:1994}:
\begin{equation}\label{FQ2}
F_Q(\xi)=2\sum_{m,n}\frac{|\langle\psi_m|\partial_\xi\rho|\psi_n\rangle|^2}{p_m+p_n}\;.
\end{equation}
Specifically, for a two-level system, the quantum state  is  written  in  terms of the Bloch sphere representation. Then, by inserting Eq.~(\ref{Blochv}) into Eq.~(\ref{FQ2}), we have~\cite{Zhong:2013}
\begin{equation}\label{Fq3}
F_Q(\xi)=\left\{\begin{aligned}&|\partial_\xi\bm{\omega}|^2
+\frac{(\bm{\omega}\cdot\partial_\xi\bm{\omega})^2}{1-|\bm{\omega}|^2},&\quad|\bm{\omega}|<1\;;\\
&|\partial_\xi\bm{\omega}|^2,&\quad|{\bm{\omega}}|=1\;.
\end{aligned}\right.
\end{equation}
Note that the Bloch vector generally  satisfies $|\bm{\omega}|<1$.
Thus, once
exact evolution of the Bloch vector is determined, the QFI for the  detector can be obtained straightforwardly
from Eq.~(\ref{Fq3}), and therefore a parameter estimation can be
performed.

For the longtime (asymptotic) limit $\tau\rightarrow\infty$, we can define the asymptotic QFI  as
\begin{equation}\label{F-asym}
F_Q^{\rm{asy}}(\xi):=\lim_{\tau\to\infty}{F_Q(\xi)}\;.
\end{equation}
From Eq.~(\ref{bloch-re}), it is easy to verify that
\begin{equation}\label{F-asym2}
F_Q^{\rm{asy}}(\xi)=\lim_{\tau\rightarrow\infty}F_C(\xi)=\frac{1}{1-(B/A)^2}\Big[\frac{\partial({B}/{A})}{\partial\xi}\Big]^2\;.
\end{equation}

Moreover, when setting $\theta=k\pi$ for $k\in{\bm{Z}}$, the QFI exactly equals to the classical Fisher information, i.e., $F_Q(\xi)\big|_{\theta=k\pi}=F_C(\xi)$. Notably  when choosing the weight parameter as the parameter of interest, it follows from Eq.~(\ref{Fq3}) and the solution of Bloch vector~(\ref{bloch-re}) that the QFI is an even function with respect to the weight parameter $\theta$. In the following sections, we employ the exact temporal evolution of the detector system to perform parameter estimations for an accelerated detector in various scenarios.

\section{Parameter estimations in various constant acceleration scenarios}
Before delving into the problem of parameter estimation  for accelerated detectors, we  first introduce the common families of constant-acceleration motions. Beyond  the most well-known trajectory of linear acceleration motion for a pointlike detector, four additional classes of stationary, constant-acceleration motions in Minkowski spacetime have been identified~\cite{Letaw:1980,Good:2020}: cusped, catenary, circular, and helical motions.

For convenience,  these acceleration trajectories can be characterized using three geometric parameters, i.e.,  the curvature $a$, which represents the magnitude of proper acceleration, the first torsion $b$, associated with the proper angular velocity in a given tetrad frame,
and the second torsion (hypertorsion) $\nu$, which accounts for additional rotational effects~\cite{Letaw:1980,Bozanic:2023}. More specifically,
\begin{equation}\label{five-s}
\left\{\begin{aligned}&\text{ linear:}~a\neq0,~b=\nu=0;\\
&\text{catenary:}~a>b,~\nu=0;\\
&\text{ cusped:}~a=b,~\nu=0;\\
&\text{circular:}~a<b,~\nu=0;\\
&\text{ helix: }~\forall{b}, a\neq0,\nu\neq0.
\end{aligned} \right.
\end{equation}
In (3+1)-dimensional Minkowski spacetime, the corresponding stationary acceleration trajectories can be expressed as follows:
\begin{equation}\label{Five-t}
x(\tau)=\left\{\begin{aligned}&\Big(\frac{\sinh(a\tau)}{a},~\frac{\cosh(a\tau)}{a},~ 0,~ 0\Big),&\text{linear}\\
&\Big(\frac{a\sinh(\sqrt{a^2-b^2}\tau)}{a^2-b^2},~\frac{a\cosh(\sqrt{a^2-b^2} \tau)}{a^2-b^2},~\frac{b\tau}{\sqrt{a^2-b^2}},~0\Big),&\text{catenary}\\
&\Big(\tau+\frac{a^2\tau^3}{6},~ \frac{a\tau^2}{2}, ~\frac{a^2\tau^3}{6},~ 0\Big),&\text{cusped}\\
& \Big(\frac{b\tau}{\sqrt{b^{2}-a^{2}}},~\frac{a\cos(\sqrt{b^{2}-a^{2}}\tau)}{b^{2}-a^{2}},~\frac{a\sin(\sqrt{b^{2}-a^{2}}\tau)}{b^{2}-a^{2}},~0\Big),&\text{circular}\\
&\Big(\frac{\mathcal{P}\sinh(\Gamma_+\tau)}{\Gamma_+},~
\frac{\mathcal{P}\cosh(\Gamma_+\tau)}{\Gamma_+},~
\frac{\mathcal{Q}\cos(\Gamma_-\tau)}{\Gamma_-},~
\frac{\mathcal{Q}\sin(\Gamma_-\tau)}{\Gamma_-}\Big),&\text{helix}\;,
        \end{aligned} \right.
\end{equation}
where $\mathcal{P}:=\Xi/\Gamma, \mathcal{Q}:=ab/(\Xi\Gamma)$ with
\begin{equation}
\begin{aligned}
&\Xi^{2}:=\frac{1}{2}(\Gamma^{2}+a^{2}+b^{2}+\nu^{2}),
~\Gamma^{2}:=\Gamma_{+}^{2}+\Gamma_{-}^{2},\\&\Gamma_{\pm}^{2}:=\sqrt{\mathcal{A}^{2}+\mathcal{B}^{2}}\pm\mathcal{A}, \\
&\mathcal{A}:=\frac{1}{2}(a^{2}-b^{2}-\nu^{2}),~ \mathcal{B}:=a\nu.
\end{aligned}
\end{equation}
 In particular, when $\nu=0$,  the Wightman functions for the first four trajectories in Eq.~(\ref{Five-t})
 can be given in  the simple $i\epsilon-$form~\cite{Bozanic:2023},
\begin{equation}\label{wigh-all}
W_{\nu=0}(\Delta\tau)=-\frac{1}{4\pi^{2}}\Bigg[-\frac{\bar{b}^{2}\Delta\tau^{2}}{1-\bar{b}^{2}}
+\frac{4\sinh^{2}({a\Delta\tau\sqrt{1-\bar{b}^2}}/{2})}{(1-\bar{b}^{2})^{2}a^{2}}-{\rm{sgn}}(\Delta\tau)i\epsilon\Bigg]^{-1}\;,
\end{equation}
where $\bar{b}:=b/a$ and  ${\rm{sgn}}(x)$ denotes the sign function.

When $\nu\neq0$, the Wightman function for helix  trajectory is given by
\begin{equation}\label{wigh-hel}
W_{\rm{hel}}(\Delta\tau)=-\frac{1}{4\pi^{2}}\Bigg[\frac{4\mathcal{P}^2}{\Gamma^2_+}\sinh^2\Big(\frac{\Gamma_+\Delta\tau}{2}\Big)
-\frac{4\mathcal{Q}^2}{\Gamma^2_-}\sin^2\Big(\frac{\Gamma_-\Delta\tau}{2}\Big)-{\rm{sgn}}(\Delta\tau)i\epsilon\Bigg]^{-1}\;.
\end{equation}
 This function reduces to  Eq.~(\ref{wigh-all})  in the limit of $\nu\rightarrow0$, showing that the helical trajectory generalizes the other four cases. For small $\nu$,  the dynamics and metrological behavior of detectors following a helix are thus qualitatively similar to those in the other four trajectories, with the parameter $\bar{b}$ determining which motion the helix most closely resembles. For large $\nu$,  the Wightman function exhibits increasing similarity to that of linear acceleration.  In the limit  $\nu\gg1$, one finds  $\mathcal{P}^2\sim1,~\mathcal{Q}^2\sim{a^4\bar{b}^2/\nu^4},~\Gamma^2_+\sim{a^2}$, and $~\Gamma^2_-\sim{\nu^2}$, leading to the approximation:
 \begin{equation}
 W_{\rm{hel}}(\Delta\tau)\approx W_{\nu=0}(\Delta\tau)\Big|_{\bar{b}=0}=-\frac{1}{16\pi^{2}}\Big[\frac{\sinh^2(a\Delta\tau/2)}{a^2}-{\rm{sgn}}(\Delta\tau)i\epsilon\Big]^{-1}\;.
 \end{equation}
 Thus, for large $\nu$
 the behavior of detectors on a helical trajectory can be analyzed through its close analogy with linear acceleration. In this sense, once the dynamics and metrology of the other four trajectories are well understood, one can gain a comprehensive understanding of the helical case.

According to Eq.~(\ref{five-s}),  parameter $\bar{b}$ for the case of $\nu=0$ can   specify the four acceleration motions:
 linear ($\bar{b}=0$), catenary
($0 <\bar{b}< 1$), cusped ($\bar{b}=1$), and circular ($\bar{b}>1$). Substituting Eq.~(\ref{wigh-all}) into Eq.~(\ref{gomega}), we have
\begin{align}\label{gomega-2}
\mathcal{G}(\Omega)&=-\frac{1}{4\pi^{2}}\int_{-\infty}^{\infty} \frac{e^{i\Omega\Delta\tau}d\Delta\tau}{-\frac{\bar{b}^{2}\Delta\tau^{2}}{1-\bar{b}^{2}}
+\frac{4\sinh^{2}({a\Delta\tau\sqrt{1-\bar{b}^2}}/{2})}{(1-\bar{b}^{2})^{2}a^{2}}-{\rm{sgn}}(\Delta\tau)i\epsilon}\nonumber\\
&=-\frac{1}{4\pi^{2}}\Bigg[\int_{-\infty}^{\infty} \frac{e^{i\Omega\Delta\tau}d\Delta\tau}{-\frac{\bar{b}^{2}\Delta\tau^{2}}{1-\bar{b}^{2}}
+\frac{4\sinh^{2}({a\Delta\tau\sqrt{1-\bar{b}^2}}/{2})}{(1-\bar{b}^{2})^{2}a^{2}}-{\rm{sgn}}(\Delta\tau)i\epsilon}
-\int_{-\infty}^{\infty}\frac{e^{i\Omega\Delta\tau}d\Delta\tau}{(\Delta\tau-i\epsilon)^{2}}+\int_{-\infty}^{\infty}\frac{e^{i\Omega\Delta\tau}d\Delta\tau}{(\Delta\tau-i\epsilon)^{2}}\Bigg]
\nonumber\\
&=
\frac{1}{4\pi^{2}}\int_{-\infty}^{\infty}
\frac{\big[2-a^{2}\Delta\tau^{2}(-1+\bar{b}^{2})-2\cosh(a\Delta\tau\sqrt{1-\bar{b}^{2}})\big]e^{i\Omega\Delta\tau}d\Delta\tau}{
\big[2-a^2\bar{b}^2\Delta\tau^2(-1+\bar{b}^2)-2\cosh(a\Delta\tau\sqrt{1-\bar{b}^2})\big]\Delta\tau^{2}}-\frac{1}{4\pi^{2}}\int_{-\infty}^{\infty}\frac{e^{i\Omega\Delta\tau}d\Delta \tau}{(\Delta\tau-i\epsilon)^{2}}\nonumber\\
&=
\frac{1}{2\pi^{2}}\int_{0}^{\infty}
\frac{2-a^{2}\Delta\tau^{2}(-1+\bar{b}^{2})-2\cosh(a\Delta\tau\sqrt{1-\bar{b}^{2}})}{
2-a^2\bar{b}^2\Delta\tau^2(-1+\bar{b}^2)-2\cosh(a\Delta\tau\sqrt{1-\bar{b}^2})}\frac{\cos(\Omega\Delta\tau)}{\Delta\tau^{2}}d\Delta\tau+\frac{\Omega}{2\pi}\Theta(\Omega)\;
\end{align}
with  $\Theta(x)$ denoting the Heaviside step function.
 It is worth noting that  the added and subtracted term in the second line of Eq.~(\ref{gomega-2}) precisely corresponds to the Fourier transform of the Wightman function for  an inertial trajectory. This  technique  is commonly used in calculating the transition probabilities for an Unruh-DeWitt detectors in quantum field theory (see Refs.~\cite{Louko:2006,Satz:2007,Zhjl:2020}), as it regularizes the integrand in the final line of Eq.~(\ref{gomega-2}), thereby facilitating direct numerical integration.  In particular, for $\bar{b}\neq0,1$, it is more convenient to use this form for numerical evaluation than the original expression in the first line, which involves integrable singularities requiring special handling.  Accordingly,
the coefficients in the Kossakowski matrix, $A$ and $B$, can be obtained
straightforwardly by inserting Eq.~(\ref{gomega-2}) into
Eq.~(\ref{ABcs}). However, the effective energy gap
$\tilde{\Omega}$, which involves the Hilbert transform
$\mathcal{K}(\Omega)$, needs a suitable renormalization procedure due to the
nonrelativistic disposal of the Unruh-DeWitt
model\cite{Bethe:1947,Milonni:1994,Breuer:2002,Benatti:2004}. In the
weak-coupling assumption ($\lambda\rightarrow0$), one can actually
omit  the Lamb shift terms (relevant to the Hilbert transform) of
the effective Hamiltonian~(\ref{heff-r}) in parameter estimation owing
to their  small contributions to the detector's dynamical evolution.

\subsection{The quantum Fisher information for the acceleration parameter}
In the following, we will explore the precision limits of parameter estimation for accelerated detectors across different acceleration scenarios. Here, the parameter of interest, $\xi$, can be  specified as the acceleration parameter $a$. For the linear and cusped cases,  the Fourier transforms of the Wightman functions can be obtained analytically by using residue theorem,
\begin{equation}\label{gomega-app1}
\mathcal{G}(\Omega)=\left\{\begin{aligned}
&\frac{\Omega}{4\pi}\big[1+\coth({\pi\Omega}/{a})\big]\;,~~&\bar{b}=0\;;
\\&\frac{\sqrt{3}a}{24\pi}{e}^{-2\sqrt{3}|\Omega|/a}+\frac{\Omega}{2\pi}\Theta(\Omega)\;,~~&\bar{b}=1\;.
\end{aligned}\right.
\end{equation}
Additionally, for the catenary and circular cases,
according to Eq.~(\ref{gomega-2}), the Fourier  transforms of the Wightman function, $\mathcal{G}(\Omega)$,  can be approximated in certain special cases
\begin{equation}\label{gomega-app2}
\mathcal{G}(\Omega)\approx\left\{\begin{aligned}&\frac{\Omega}{4\pi}\big[1+\coth({\pi\Omega}/{a})\big]\;,~~&{a}/\Omega\ll1,0<\bar{b}\ll1\;;\\
&\frac{\sqrt{3}a}{24\pi}{e}^{-2\sqrt{3}|\Omega|/a}+\frac{\Omega}{2\pi}\Theta(\Omega)\;,~~&{a}/\Omega\ll1, 1<\bar{b}<\Omega/a\;;\\&
\frac{\Omega}{4\pi}(1-\bar{b}^2)\big[1+\coth({\pi\Omega}/{a})\big]\;,~~&a/\Omega\gg1,0<\bar{b}\ll1\;;
\\&\frac{13a}{24\bar{b}\pi^2}-\frac{|\Omega|}{4\bar{b}^2\pi}+\frac{\Omega}{2\pi}\Theta(\Omega)\;,~~&a/\Omega\gg1,1\ll\bar{b}<a/\Omega\;.
\end{aligned}\right.
\end{equation}
For convenience, we can rescale the time variable
$\tau\mapsto\tilde{\tau}\equiv\lambda^2\tau$ in~(\ref{bloch-re}) and continue  referring to $\tilde{\tau}$ as
$\tau$ in the following analysis. This transformation ensures that the evolution equation
 attains a well-defined limit as
$\lambda \rightarrow 0$ .

Using Eqs.~(\ref{bloch-re})~(\ref{Fq3})~(\ref{gomega-app1}) and~(\ref{gomega-app2}),  we approximate the QFI  for vanishingly small acceleration ($a/\Omega\ll1, a\tau\ll1$) and small evolution time~\footnote{Here, the small evolution time refers to the rescaled variable  $\tilde{\tau} \Omega=\lambda^2\tau\Omega\ll1$, where $\tau$  denotes the physical (nonrescaled) time.  As argued in Refs.~\cite{Moustos:2017,Kaplanek:2020}, the Markov approximation remains valid for sufficiently late times. Under the weak-coupling assumption ($\lambda\rightarrow0$), it is possible to satisfy both  $\tilde{\tau} \Omega\ll1$ and
 $\tau\Omega>1$,  thus justifying the use of the Markov approximation in this regime.}
(${\tau}\ll1/\Omega$)
\begin{equation}\label{fa-a-small}
F_Q(a)\approx\left\{\begin{aligned}
&\frac{\Omega^3\pi\tau{e}^{-4\pi\Omega/a}}{8 a^4}
\cdot\frac{[3+\cos(2 \theta)]^2}{\cos^4(\theta/2)+e^{-2\pi\Omega/{a}}\sin^4(\theta/2)},~~&0\leq\bar{b}\ll1\;;\\
& \frac{3\Omega^2\tau{e}^{-\frac{4\sqrt{3}\Omega}{a}}}{8 a^2 \pi}
\cdot\frac{[3+\cos(2 \theta)]^2}{48 \Omega
\cos^4(\frac{\theta}{2})+\sqrt{3} a e^{-\frac{2 \sqrt{3}\Omega}{{a}}}[3+\cos(2
\theta)]}, &1\leq\bar{b}<\Omega/a\;.
\end{aligned} \right.
\end{equation}
This result indicates that  the QFI for the acceleration parameter, $F_Q(a)$,  increases with time for short evolution durations, and  the weight parameter plays a significant role in determining the Fisher information.   Note that in the limit  $a\rightarrow0$, the  QFI approaches zero for all trajectories, regardless of the value of the acceleration-scenario parameter  $\bar{b}$.

For large accelerations and sufficiently long evolution times
($a/\Omega\gg1$, $a\tau\gg1$), the  QFI can be approximated as
\begin{equation}\label{fa-a-large}
F_Q(a)\approx\left\{\begin{aligned}
&\frac{\pi^2\Omega^2}{a^4(1-\bar{b}^2)}+\frac{(1-\bar{b}^2)^3\tau^2\sin^2{\theta}}{16\pi^4}e^{{-a\tau}/(2\pi^2)},~~&0\leq\bar{b}\ll1\;;\\
& \frac{12\Omega^2}{a^4}+\frac{\tau^2\sin^2{\theta}}{192\pi^2}e^{{-a\tau}/(4\sqrt{3}\pi)},~~&\bar{b}=1\;;\\
&\frac{36\bar{b}^2\pi^2\Omega^2}{169a^4}+\frac{169\tau^2\sin^2\theta}{576\bar{b}^2\pi^{4}}e^{{-{13a\tau}/(12\bar{b}\pi^{2})}},&1\ll\bar{b}<a/\Omega\;.
\end{aligned} \right.
\end{equation}
These expressions highlight the significant influence of $\bar{b}$ on the QFI for the acceleration parameter in the large acceleration regime. Over long evolution times,
 the QFI for the acceleration parameter asymptotically converges to a nonnegative value,  which remains independent of the weight parameter.
Substituting  Eqs.~(\ref{ABcs})~(\ref{gomega-app1}) and~(\ref{gomega-app2}) into Eq.~(\ref{F-asym2}), we obtain the asymptotic QFI for vanishingly small accelerations
\begin{equation}\label{Fqas-small}
F_Q^{\rm{asy}}(a)\approx\left\{\begin{aligned}&\frac{4\pi^2\Omega^2}{a^4}e^{-2\pi\Omega/a}\;,~~&a/\Omega\ll1, 0\leq\bar{b}\ll1\;;
\\&\frac{\sqrt{3}\Omega}{a^3}e^{-2\sqrt{3}\Omega/a}\;,~~&a/\Omega\ll1,1\leq\bar{b}<\Omega/a\;.
\end{aligned}\right.
\end{equation}
Similarly, for large accelerations, we have
\begin{equation}\label{Fqas-larg}
F_Q^{\rm{asy}}(a)\approx\left\{\begin{aligned}
&\frac{\pi^2\Omega^2}{a^4(1-\bar{b}^2)}\;,~~&a/\Omega\gg1,0\leq\bar{b}\ll1\;;
\\&\frac{12\Omega^2}{a^4}\;,~~&a/\Omega\gg1,\bar{b}=1\;;\\&
\frac{36\bar{b}^2\pi^2\Omega^2}{169a^4}\;,~~&a/\Omega\gg1,1\ll\bar{b}<a/\Omega\;.
\end{aligned}\right.
\end{equation}

From the first line of Eq.~(\ref{Fqas-larg}), we can clearly see that the asymptotic QFI increases as $\bar{b}$  increases from 0 to 1, implying that the QFI for the catenary case (small $\bar{b}$)  exceeds that of the linear case ($\bar{b}=0$).  Comparing the second line to the first, the QFI in the cusped case ($\bar{b}=1$) is larger than that in the small-$\bar{b}$  catenary regime.
Furthermore, we note that in circular motion, the linear velocity is proportional to $1/\bar{b}$, so the condition  $\bar{b}\gg1$ corresponds to the low-velocity regime. The third line of Eq.~(\ref{Fqas-larg}) then shows that the QFI in the circular case becomes the largest among all scenarios when $\bar{b}>2.4$.
Thus, for  $a/\Omega\gg1$,  we conclude that the asymptotic QFI for the acceleration parameter follows the ascending order:   linear $<$ catenary(small $\bar{b}$) $<$  cusped $<$ circular case (low velocity).
This ordering provides valuable insight into which acceleration scenarios are optimal for quantum metrology applications.

To quantify the precision of estimating the acceleration parameter in different acceleration scenarios, we numerically compute the corresponding QFI for the acceleration parameter  by evaluating the integral~(\ref{gomega-2}).
\begin{figure}[!ht]
  \centering
\subfloat[$a/\Omega=0.5,\theta=\pi/4$]{\label{Fat1}\includegraphics[width=0.47\linewidth]{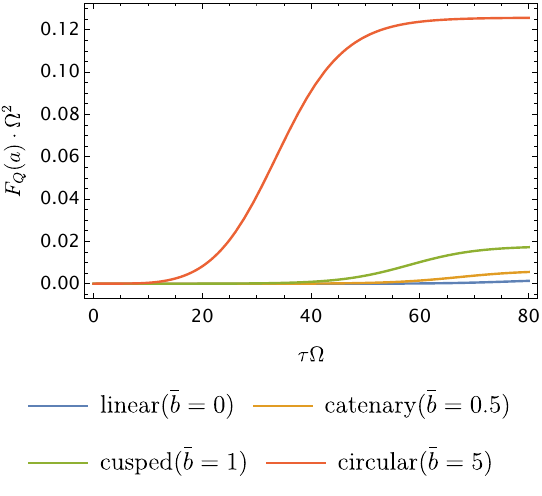}}~
\subfloat[$a/\Omega=10,\theta=\pi/4$]{\label{Fat2}\includegraphics[width=0.47\linewidth]{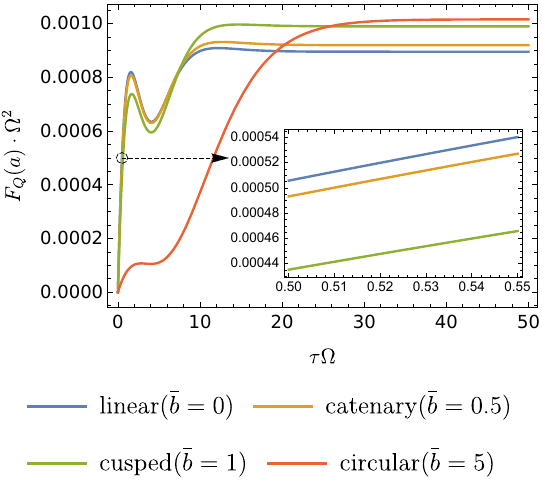}}\\
\subfloat[$a/\Omega=0.5,\theta=\pi$]{\label{Fat3}\includegraphics[width=0.47\linewidth]{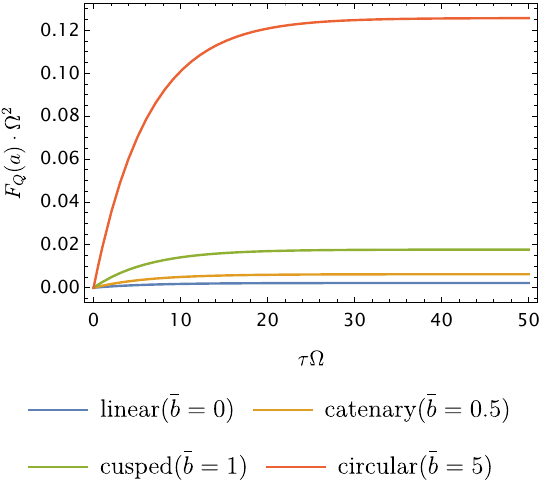}}~
 \subfloat[$a/\Omega=10,\theta=\pi$]{\label{Fat4}\includegraphics[width=0.47\linewidth]{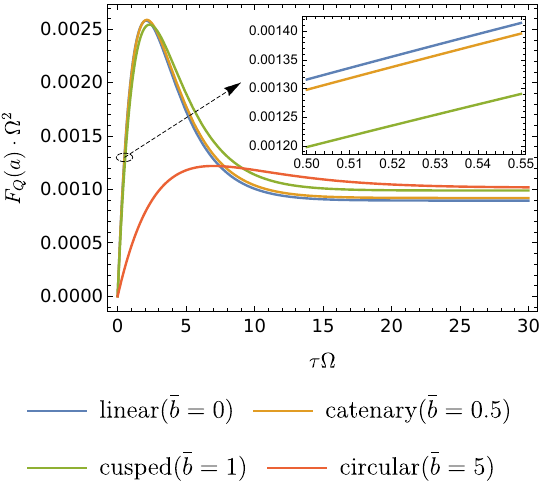}}
  \caption{The QFI for the acceleration parameter is plotted as a function of evolution time for various acceleration and  weight parameters.  For convenience,
all relevant physical quantities  are expressed in units of the detector's energy gap $\Omega$.}
  \label{fig-Fat}
\end{figure}
Figure~(\ref{fig-Fat}) illustrates the temporal evolution of the QFI for the acceleration parameter in the four different acceleration scenarios. For small accelerations ($a/\Omega<1$),  we  observe  that  the QFI, $F_Q(a)$, generally increases over time and eventually saturates to a finite nonnegative value. This implies that as the detector evolves, its sensitivity to variations in the acceleration parameter improves. Consequently, a sufficiently long evolution time allows for optimal estimation of the acceleration parameter, a conclusion consistent with that of Ref.~\cite{Tian:2015}. However, for large accelerations ($a/\Omega\gg1$), the QFI exhibits a different behavior. Initially, it grows rapidly, then oscillates, and ultimately converges to a positive asymptotic value. Notably, in certain cases [e.g., Fig.~(\ref{Fat4})],  the QFI, across all acceleration scenarios, including  the linear acceleration, reaches its maximum value  at a finite evolution time for specific values of $\theta$. This stands in sharp contrast with the main conclusion of Ref.~\cite{Tian:2015}, which asserted  the optimal precision for estimating the acceleration parameter in the linear acceleration case is achieved only after an indefinitely long evolution time. It is worth noting that the range of acceleration parameters considered in Ref.~\cite{Tian:2015} is limited to the relatively low-acceleration regime. This explains why the behavior of QFI at large accelerations, as revealed in the present work, sharply contrasts with the conclusions of Ref.~\cite{Tian:2015}.

By comparing different acceleration scenarios, we observe the following trends: for small accelerations ($a/\Omega<1$), the ascending rank ordering of $F_Q(a)$ consistently follows the sequence from linear to catenary to cusped and then to circular, while for
large accelerations ($a/\Omega\gg1$), initially  the ascending rank ordering   is circular to cusped to catenary and then to linear scenario, but for sufficiently long evolution times, this ranking is reversed.  Therefore, the circular acceleration scenario provides the highest estimation precision when the acceleration is much smaller than the detector's energy gap or when the evolution time is sufficiently long if the acceleration is large.

\begin{figure}[!ht]
  \centering
  \subfloat[$\theta=\pi/4,\tau\Omega=1$]{\label{Faa1}\includegraphics[width=0.45\linewidth]{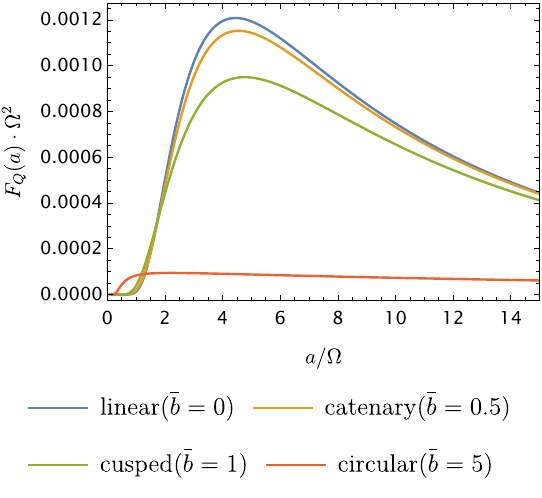}}~
  \subfloat[$\theta=\pi,\tau\Omega=1$]{\label{Faa2}\includegraphics[width=0.45\linewidth]{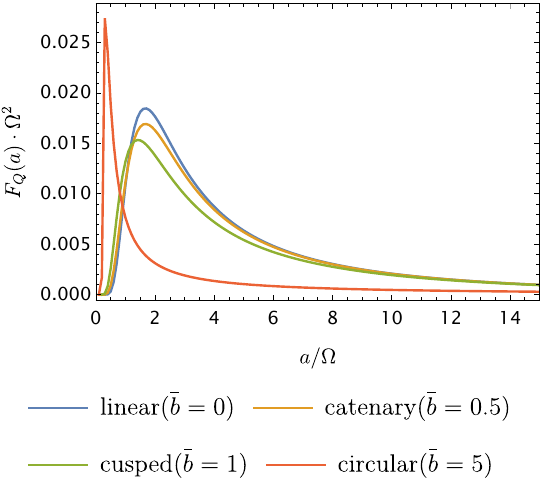}}
  \caption{The QFI for the acceleration parameter versus $a/\Omega$ with $\theta=\pi/4,\tau\Omega=1$ in (a), and with $\theta=\pi,\tau\Omega=1$ in (b).} \label{fig-Faa}
\end{figure}

\begin{figure}[!ht]
  \centering
  \subfloat[$a/\Omega=0.5,\tau\Omega=1$]{\label{Fath1}\includegraphics[width=0.33\linewidth]{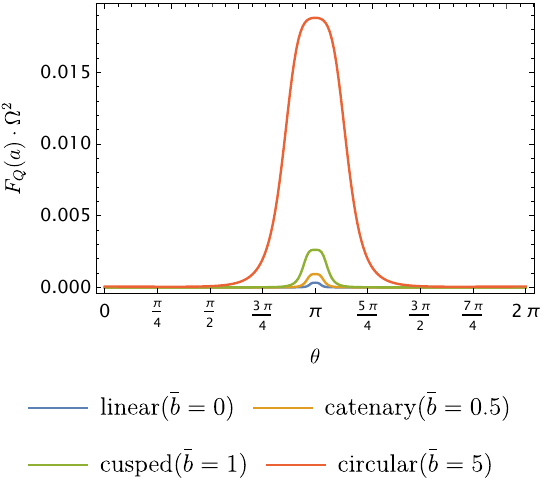}}~
  \subfloat[$a/\Omega=1,\tau\Omega=1$]{\label{Fath2}\includegraphics[width=0.33\linewidth]{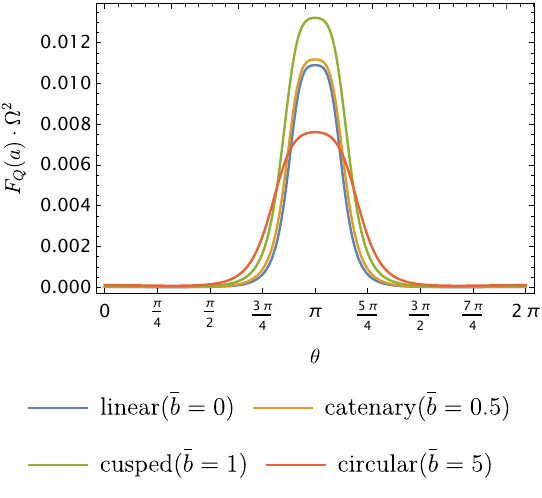}}~
  \subfloat[$a/\Omega=5,\tau\Omega=1$]{\label{Fath3}\includegraphics[width=0.33\linewidth]{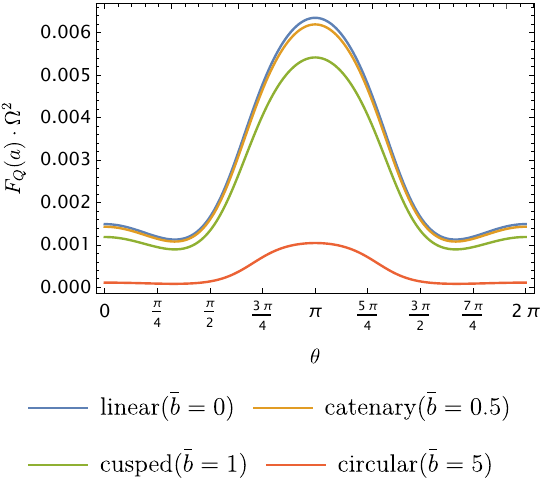}}
  \caption{The QFI for the acceleration parameter versus $\theta$ with $a/\Omega=0.5$ in (a), $a/\Omega=1$ in (b) and $a/\Omega=5$ in (c). Here, we assume $\tau\Omega=1$ for all plots.} \label{fig-Fath}
\end{figure}

To further examine the influence of acceleration $a$ and the weight parameter $\theta$  on the QFI, we respectively
plot $F_Q(a)\Omega^2$ versus $a/\Omega$ and $\theta$ for a finite evolution time in Figs.~(\ref{fig-Faa}) and~(\ref{fig-Fath}).  From Fig.~(\ref{fig-Faa}), we observe:  the QFI exhibits a peak at certain nonzero values of
$a/\Omega$,  indicating an optimal acceleration value for parameter estimation. The linear acceleration scenario seems to have relatively large values of  QFI as  $a/\Omega$ is not too small. In Fig.~(\ref{fig-Fath}), we find that the QFI for the acceleration parameter is maximized at $\theta=(2k+1)\pi$ for $k\in{\bf{Z}}$, i.e., when the detector initially is in its ground state.  In general, increasing the acceleration tends to reduce the peak value of QFI. Moreover, for small accelerations and finite evolution time, the circular trajectory yields the highest QFI peak as a function of $\theta$, whereas for large accelerations, the linear case becomes optimal.

\begin{figure}[!ht]
  \centering
 \includegraphics[width=0.65\linewidth]{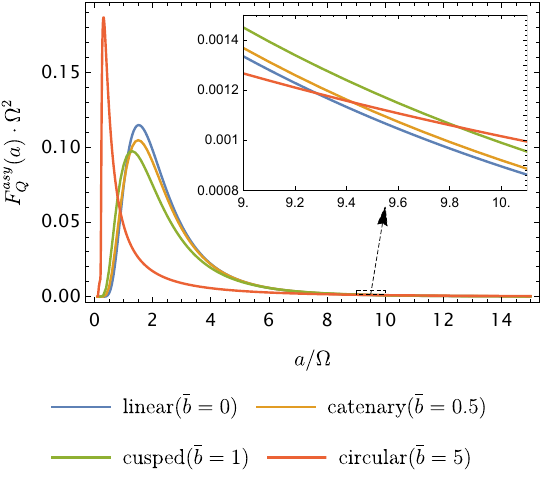}
 \caption{The asymptotic QFI for the acceleration parameter is plotted as a function of $a/\Omega$ .}
  \label{fig-Fasym}
\end{figure}

In the limit of $\tau\rightarrow\infty$, the QFI approaches its  asymptotic value, $F_Q^{\rm{asy}}(a)$, which,
 according to Eq.~(\ref{F-asym2}), depends only on acceleration and its specific trajectory but not on the weight parameter $\theta$.
In Fig.~(\ref{fig-Fasym}), we plot the asymptotic QFI as a function of the acceleration parameter for different trajectories.
From Fig.~(\ref{fig-Fasym}),  we conclude: For accelerations comparable to the energy gap ($a\sim\Omega$), the asymptotic QFI exhibits a peak in all scenarios except the circular case. The circular scenario attains the largest peak at a smaller value of  $a/\Omega$  compared to the other scenarios. The ascending order of the peaks  in the remaining three acceleration scenarios follows cusped to catenary to linear.
Remarkably, for both  quite small ($a/\Omega<1 $) and extremely large ($a/\Omega\gg1 $) accelerations, the
asymptotic QFI follows an ascending rank ordering of linear, catenary, cusped, and then to circular cases, which agrees with the earlier conclusion  analytically derived from Eq.~(\ref{Fqas-larg}).

 Overall, these results provide deeper an insight into the estimation of the acceleration parameter in different relativistic motion scenarios, highlighting the importance of trajectory selection in quantum metrology applications.

 \subsection{The quantum Fisher information for the weight parameter }
 In this section, we investigate the parameter estimation for the weight parameter $\theta$ and explore which acceleration profiles introduce the most noise into its quantum measurement.
 Before proceeding, let us note that  in quantum theory, weight parameters typically refer to  the coefficients that characterize a superposition of states. The precise estimation of these parameters  is crucial for optimizing the performance of quantum algorithms, enhancing the stability of quantum devices, and enabling more efficient quantum control strategies. As such, the estimation of weight parameters plays a central role in quantum computation and quantum information processing. The estimation of the weight parameter is inevitably affected by acceleration-induced thermal noise, which can degrade the performance of quantum algorithms and hinder the development of high-efficiency quantum control strategies. In the following, we analyze how different acceleration profiles influence the estimation precision of the weight parameter, and identify which scenarios yield optimal precision.

 Let the parameter of interest,  $\xi$,  be the weight parameter $\theta$.   Its QFI~(\ref{Fq3}) can be approximated in certain special  cases. Specifically, based on the approximation~(\ref{gomega-2}), when the acceleration is vanishingly small compared to the energy gap ($a/\Omega\ll1, a\tau\ll1$),  the QFI for short evolution  times (${\tau}\ll1/\Omega$) can be written as
 \begin{equation}\label{ftheta-a-small}
 F_Q(\theta)\approx\left\{\begin{aligned}
&e^{-\Omega\tau/(2\pi)} \Big[1-\frac{\sin^2\theta}{1 + e^{2\pi\Omega/a} \cos^4(\theta/2)}\Big],~~&0\leq\bar{b}\ll1\;;\\
&{e}^{{-\Omega\tau}/(2\pi)} \Big[1-\frac{a\Omega\tau\sin^2\theta}{8\sqrt{3}e^{{2 \sqrt{3}\Omega}/a}\Omega\pi \cos^4(\theta/2)-2a \pi\cos\theta}\Big], &1\leq\bar{b}<\Omega/a\;.
\end{aligned} \right.
\end{equation}
This expression shows that the QFI decreases over time, which aligns with the intuitive expectation that quantum information about the weight parameter is gradually degraded by acceleration-induced thermal noise during the detector's thermalization process.

For large accelerations  and not too small evolution time
($a/\Omega\gg1$, $a\tau\gg1$), the QFI can be approximated as
\begin{equation}\label{ftheta-a-large}
 F_Q(\theta)\approx\left\{\begin{aligned}
&e^{-{a(1-\bar{b}^2)^{3/2}\tau}/(2 \pi^2)}\cos^{2}\theta+e^{-{a(1-\bar{b}^2)^{3/2}\tau}/{\pi^2}}(1+\cos^2\theta)\sin^2\theta,~&0\leq\bar{b}\ll1\;;\\
&e^{-{a\tau}/(4\sqrt{3}\pi)}\cos^{2}\theta+e^{-{a\tau}/(2\sqrt{3}\pi)}(1+\cos^2\theta)\sin^2\theta, ~&\bar{b}=1\;;\\
&e^{-13a\tau/(12\bar{b}\pi^2)}\cos^2\theta+e^{-13a\tau/(6\bar{b}\pi^2)}(1+\cos^2\theta)\sin^2\theta, &1\ll\bar{b}<a/\Omega.
\end{aligned} \right.
\end{equation}
This result indicates that the parameter $\bar{b}$ has a significant impact on
 the QFI. As the evolution time increases, the QFI for the weight parameter  gradually diminishes to zero, meaning that in the longtime limit, all information about $\theta$ is effectively lost due to thermalization effects induced by acceleration.

\begin{figure}[!ht]
   \subfloat[$a/\Omega=1,\theta=\pi/4$]{\label{Ftht1}\includegraphics[width=0.47\linewidth]{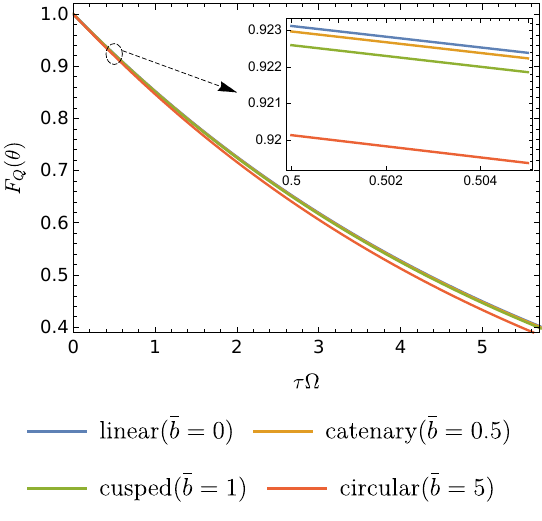}}~
\subfloat[$a/\Omega=10,\theta=\pi/4$]{\label{Ftht2}\includegraphics[width=0.47\linewidth]{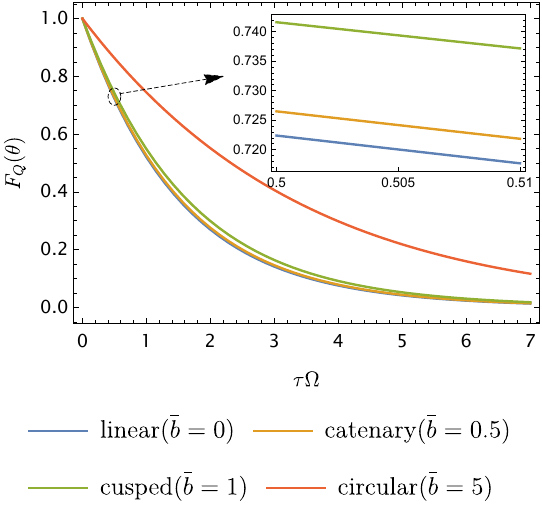}}\\
\subfloat[$a/\Omega=1,\theta=\pi$]{\label{Ftht3}\includegraphics[width=0.47\linewidth]{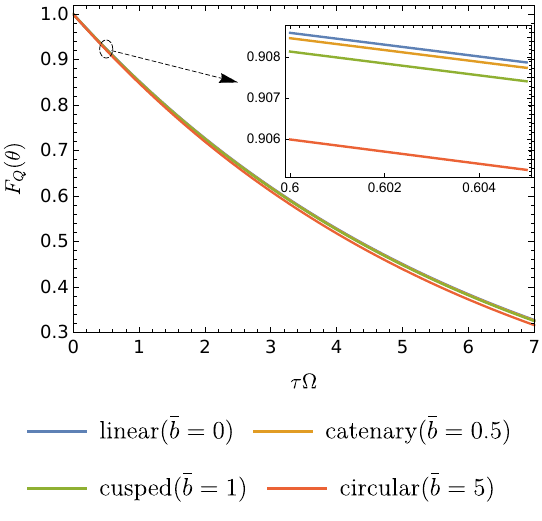}}~
 \subfloat[$a/\Omega=10,\theta=\pi$]{\label{Ftht4}\includegraphics[width=0.47\linewidth]{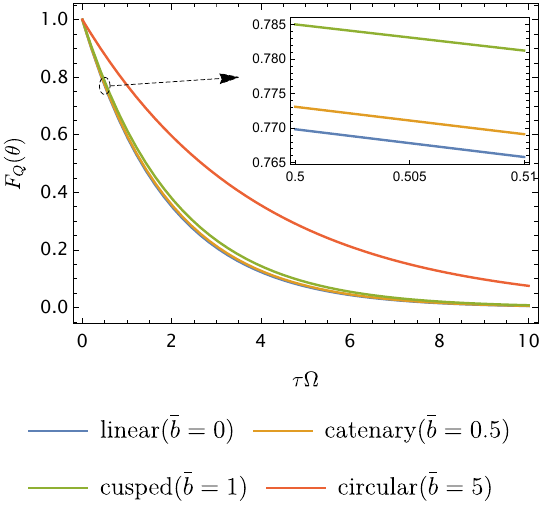}}
  \caption{The QFI for the weight parameter is plotted as a function of $\tau\Omega$ for various $a/\Omega$ and $\theta$.}\label{fig-Ftht}
 \end{figure}

To gain further insight, we numerically compute QFI, $F_Q(\theta)$,
for different acceleration profiles and plot its evolution over time in Fig.~(\ref{fig-Ftht}).
 We observe that the QFI for the weight parameter decays to zero over time for all acceleration scenarios. This is consistent with the expectation that long-term thermalization erases information about  $\theta$. When the acceleration is not significantly greater than the energy gap, the differences between various acceleration scenarios are not easily distinguishable. However, the ascending rank ordering of  $F_Q(\theta)$  from circular to cusped to catenary and then to linear cases generally holds throughout the entire evolution process. In contrast, for large accelerations ($a/\Omega\gg1$), the impact of the acceleration profile becomes more pronounced.
Notably, the ordering of QFI magnitudes  can  reverse: i.e., for not too large accelerations, the ranking follows: circular
$<$ cusped $<$ catenary $<$  linear, while for large accelerations, the ranking flips: linear $<$  catenary $<$ cusped $<$  circular. This reversal suggests that different acceleration profiles  induce varying degrees of information degradation, with circular motion preserving information better in the high-acceleration regime and linear motion doing so in the low-acceleration regime.

\begin{figure}[!ht]
  \centering
  \subfloat[$\theta=\pi/4, \tau\Omega=1$]{\label{Ftha1}\includegraphics[width=0.45\linewidth]{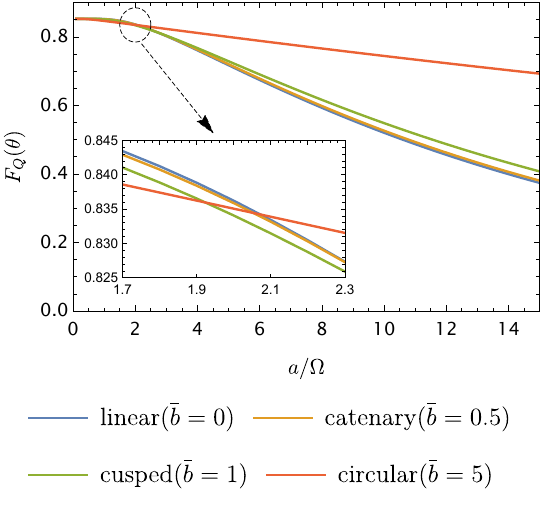}}~
  \subfloat[$\theta=\pi, \tau\Omega=1$]{\label{Ftha2}\includegraphics[width=0.45\linewidth]{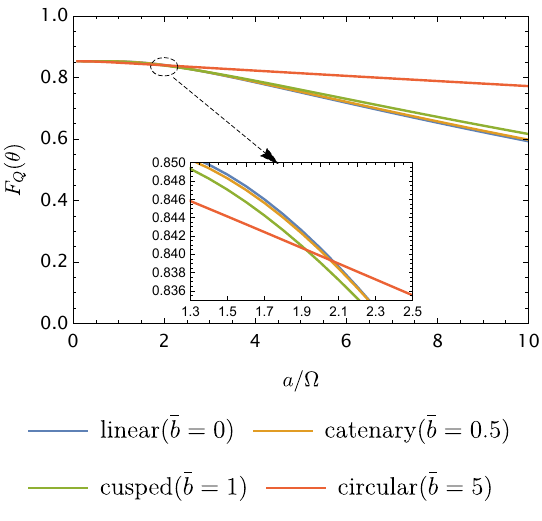}}~
  \caption{The QFI for the weight parameter is plotted as a function of $a/\Omega$ with $\theta=\pi/4,\tau\Omega=1$ in (a), and with $\theta=\pi,\tau\Omega=1$ in (b).}
  \label{fig-Ftha}
\end{figure}

\begin{figure}[!ht]
  \centering
  \subfloat[$a/\Omega=0.5, \tau\Omega=1$]{\label{Ftha}\includegraphics[width=0.33\linewidth]{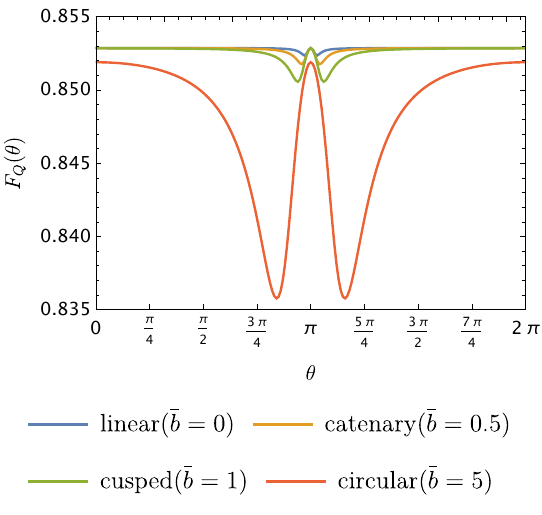}}~
  \subfloat[$a/\Omega=1, \tau\Omega=1$]{\label{Fthth}\includegraphics[width=0.33\linewidth]{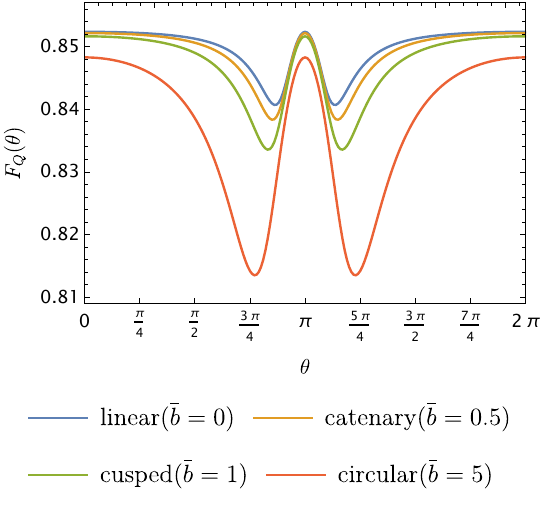}}~
  \subfloat[$a/\Omega=5, \tau\Omega=1$]{\label{Fthth}\includegraphics[width=0.33\linewidth]{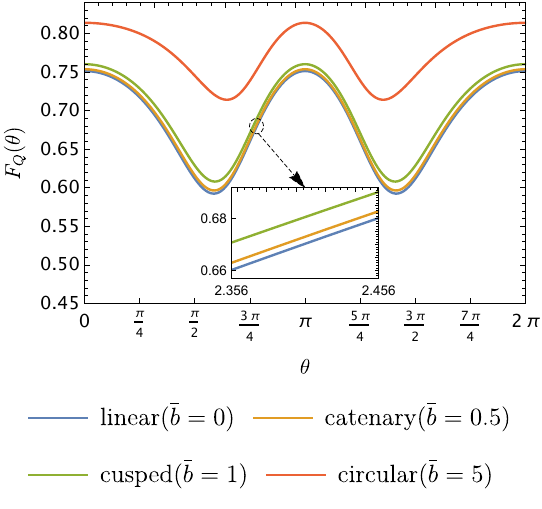}}
  \caption{The QFI for the weight parameter is plotted as a function of $\theta$ with  $a/\Omega=0.5$  in (a), $a/\Omega=1$  in (b) and  $a/\Omega=5$  in (c). Here, we have set $\tau\Omega=1$ in all plots.}
  \label{fig-Fthth}
\end{figure}

To further understand the influence of acceleration on  the QFI for the weight parameter,  we respectively plot  $F_Q(\theta)$ as  a function of both the acceleration parameter  and the weight parameter in Figs.~(\ref{fig-Ftha}) and~(\ref{fig-Fthth}). From Fig.~(\ref{fig-Ftha}), it is obvious to find  that the ascending rank ordering from circular to  cusped to catenary, and then to linear case for small accelerations will reverse at sufficiently large accelerations.
The QFI for the weight parameter  $\theta$  decreases monotonically with increasing acceleration, which  aligns with the analytical approximation~(\ref{ftheta-a-large}). In Fig.~(\ref{fig-Fthth}), the oscillatory behavior of  $F_Q(\theta)$  as a function of $\theta$  is evident, with maxima occurring at $\theta=k\pi$ ($k\in{\bm{Z}}$).
The rank ordering of QFI among the different acceleration scenarios remains  consistent across  $\theta$, reinforcing that linear acceleration is optimal in the small-acceleration regime, while circular acceleration is optimal at large accelerations.

\section{Conclusion}
\label{sec5}

In this work, we have explored the problem of parameter estimation  in
quantum metrology with an accelerated Unruh-DeWitt detector coupled
to massless scalar fields in vacuum. The detector moves along different
trajectories with constant acceleration, including linear, catenary,
cusped, and circular motions. By employing QFI as a measure of the maximal achievable precision, we have quantitatively analyzed the estimation of both the acceleration parameter
 $a$ and the weight parameter $\theta$.   Our study is based on the dynamical evolution properties of the accelerated detector within the framework of open quantum systems.

Our findings reveal that for small accelerations
($a/\Omega<1$), the QFI for acceleration parameter generally increases
monotonically with  time. This means that  the optimal
precision in estimating the acceleration parameter can be achieved by allowing
the detector to evolve for a  sufficiently long  time while moving
with a small acceleration.  However, for large accelerations
($a/\Omega\gg1$), the temporal evolution  of QFI  across all acceleration scenarios exhibits oscillatory behavior and may even reach a maximum within a finite time for an appropriately chosen weight parameter. This implies that, contrary to previous conclusions of Ref.~\cite{Tian:2015}, optimal precision in estimating the acceleration parameter can be attained at a finite evolution time rather than indefinitely long time. Regarding the ordering of QFI across different acceleration scenarios, we find that for small accelerations compared to the detector's energy gap,   the QFI follows an ascending order: linear $<$catenary $<$ cusped $<$ circular, while for large accelerations, the initial ranking order is circular $<$  cusped $<$  catenary $<$ linear, but ultimately reverses for sufficiently long evolution times.

Remarkably, the QFI for the acceleration parameter exhibits a peak value at a specific acceleration comparable
to the detector's energy gap. Moreover, for sufficiently long evolution times, the QFI  converges to a  nonnegative asymptotic value
determined by  both  the acceleration parameter and the  trajectory. The asymptotic QFI for the acceleration parameter also exhibits a peak,  with the peak value in the circular scenario being the largest among all acceleration cases. More interestingly,  for both very small and very large accelerations compared to  the detector's energy gap, the ranking of asymptotic QFI remains linear $<$catenary $<$ cusped $<$ circular (low velocity), suggesting that circular motion provides the most precise estimation of acceleration for sufficiently long evolution times.

For the weight parameter $\theta$, the QFI generally decreases monotonically over time, regardless of the acceleration scenario. Large accelerations further reduce the QFI due to thermal noise induced by the Unruh effect, leading to a complete loss of information in the longtime limit. Comparing different acceleration scenarios, we find that for small accelerations compared to the detector's energy gap, the QFI follows an ascending order: circular $<$  cusped $<$  catenary $<$ linear.  For large accelerations, this order is reversed, making circular motion the most optimal scenario for estimating the weight parameter. This result implies that, for small accelerations, linear motion introduces the least noise and thus, allows for the most accurate estimation of $\theta$. However, for large accelerations, circular motion becomes the preferred choice for minimizing measurement noise. Additionally, the initial choice of $\theta$ significantly affects  the QFI,  exhibiting oscillations with maxima occurring at $\theta =k\pi$ for $k\in{\bm{Z}}$.

Finally,  our study highlights the profound impact of different acceleration scenarios on quantum parameter estimation. The distinguishability of the QFI in different motion cases underscores the diverse evolution dynamics of accelerated detectors as quantum probes in metrology. Additionally, this work provides a foundation for further investigations into quantum metrology in relativistic settings.
A particularly intriguing direction for future research is extending our analysis to curved spacetimes. Investigating parameter estimation for an accelerated detector in the presence of gravitational effects could provide deeper insights into the interplay between quantum metrology, relativity, and fundamental physics.  Comparing these results with those in flat spacetime may offer new avenues for detecting relativistic quantum effects  in (anti-)de Sitter spacetime and even in black hole spacetimes.

%%%%%%%%%%%%%%%%%%%%

\begin{acknowledgments}
This work was supported in part by the NSFC under Grants No.~12175062 and No.~12075084, and the innovative research group of Hunan Province under Grant No. 2024JJ1006.
\end{acknowledgments}

\end{document}